\documentstyle[12pt]{article}
\input psfig
\long\def\comment#1{}
\begin{document}
\title{Collapse, Expansion, and Variable Speed of Light}
\author{Subhash Kak\\
Department of Electrical \& Computer Engineering\\
Louisiana State University\\
Baton Rouge, LA 70803-5901\\}
\maketitle

\begin{abstract}
This paper presents an information-theoretic view of how an
observer within a quantum system will 
perceive his world. 
It is argued that because of the indistinguishability of
quantum particles, a coherent state will
appear to an observer within the system like a singularity.
As superposition is lost, space appears to expand, 
although to the outsider it is merely the
collapse of the wave function.
Implications of these ideas to cosmology are considered.
The superluminal expansion of space 
provides a basis to understand inflationary cosmologies.
This
expansion may be taken to be equivalent to
a much faster speed of light during the
inflationary period. 
These ideas can be tested by checking for `higher' speed
of light from photons emitted by decohering atoms.\\


\end{abstract} 
 
\thispagestyle{empty}

\section{Introduction}
The collapse or reduction of the wave function upon measurement is perhaps
the most troubling aspect of quantum mechanics.
This collapse is supposed to occur instantaneously.
As David Bohm observed in his {\it Quantum Theory}
long ago, ``The exact region to which it collapses is not
determined by the state of the wave function before collapse...
This type of collapse of the wave function does not occur in
any classical theory.'' It is because of this that some
researchers have proposed that quantum mechanics may not
be a complete theory.
If one insisted on using classical logic or reductionist
arguments, 
this collapse leads to paradoxes such as those of the
Schr\"{o}dinger's Cat or the Wigner's Friend.

Here we reconsider the question of collapse from the
point of view of the reference frame of an 
hypothetical observer within the quantum system.
This may appear pointless until it is 
realized that the initial state of the universe
is believed to have been a pure quantum state 
described by a gigantic wave function.
Therefore, we already are observers within
an evolving system.
If this view is justified for the universe, it should
be applicable to other quantum systems as well.

The postulated wave function of the
universe at initial time has
evolved through a process that has elements
that go beyond the
framework of quantum theory (since the Schr\"{o}dinger
equation only transforms a pure state into another), or
interacted with a system outside the universe,
into a system that has classical subsystems, like the
one to which we and our measurement apparatus belongs.
In cosmology, we 
extrapolate backwards in time to estimate the nature
of time and space when the system was in a pure
quantum state. 

Quantum theory is a set of rules allowing the
computation of probabilities for the outcomes of
tests, that follow specified preparations.
In the standard interpretation of quantum mechanics,
the observation records the transition from
potentiality to actuality, but the observational
means must be described in terms of classical physics.
The observation then is a record of the interaction
between the
quantum and the classical worlds. 

Having argued that it is appropriate to consider the
frame of reference of 
an observer
within the quantum system, 
it is clear that
at Planck time from the initial condition, the universe will be one
of ceaseless creation and destruction.
But at greater time scales, one would expect
more structure and this is what we would like to
explore.

But an examination of such observations 
comes with its own problems. To
record an observation, one needs a classical
subsystem. In this preliminary study, we
postulate an imaginary observer that is 
somehow able to make measurements without
interacting with the system.
We then consider the question
of the problem of the collapse of the
wave function.
We argue that this collapse 
will appear as an expansion of the space to
an observer inside the system.

We first look at quantum information from outside.
Then the observation of a large system in a 
pure state from within is considered. This
is followed by extrapolation of these ideas to
the wave function of the universe. Lastly, we
consider the question of speed of light appearing to have
been different in the remote past.

\section{Information in a Quantum System}

Information, logarithmically proportional
to the number of possibilities (patterns), provides
a way to examine the nature of a quantum
process.
We also note that quantum theory itself has been
viewed as a theory of obtaining information about nature [1].
In a recent study [2], I argued that 
information associated with the quantum system as
it is observed from outside 
increases exponentially with the size of the measurement
apparatus.
Investigating the
information provided about a specified distributed apparatus
of $n$ units
in the measurement of a quantum state,
it was shown that, in contrast to such measurement of a classical
state, which is bounded by $\log (n + 1)$ bits, the information 
is bounded by $3.7\times  n^{\frac{1}{2}}$ bits.
This means that the use of quantum apparatus
offers an exponential gain over classical apparatus.

This unbounded information is a consequence of
the superposition at the basis of the quantum state.
It is because of this that
after the spin of a particle has been measured
to be in a particular direction, there is still a
probability $cos^2 \theta /2$ of spin in a new
direction at an angle of $\theta$ from the previous one.
One would expect that this property of unbounded
information, for which there is no parallel in the
classical world, will have an equally dramatic
analog in observations from within the quantum system.

Now, let's consider the question of information from
within the quantum system. But when we count patterns,
we cannot visualize them in classical sense
as spread over a certain physical space; we must speak of
extension for a quantum object amongst its states.

A total of $n$ indistinguishable particles
in $N$ states will be associated with
$\left( \begin{array}{c}
    N+n-1 \\ n
    \end{array} \right) $
distinct patterns if they are bosons, and with 
$\left( \begin{array}{c}
    N \\ n
    \end{array} \right) $
distinct patterns if they are fermions.
On the other hand, for distinguishable (classical) objects
the number of patterns is $ \frac{N^n}{n!}$.

Considering the number of patterns, we get most
for bosons and the least for fermions, with
classical particles somewhere in between.
For example, if $N=n=4$, the number of patterns
for fermions is 1, for classical objects is about 10, and
for bosons is 210.
This explains why 
bosons are very efficient in the transmission of information.
And it is the working of the Exclusion principle, at the basis
of the behavior of fermions, that explains the
stability of matter.

The information capacity of a system of mass $m$ has
been estimated variously using dimensional arguments
or considering the energy expended to store or
erase a bit of information [3, 4] in the presence of
the background cosmic radiation of $2.7^\circ$ K.
Since one bit of information requires $kT ln 2$,
where $k$ is the Boltzmann constant, one can,
by using the mass-energy of the universe, compute
its total information processing capacity per unit
time.

The information capacity of an electron is
about $ 3 \times 10^9$ bits per second,
while 
the total information
capacity of the universe is about $10^{98}$,
or say about $10^{100}$, bits per second.
These capacity numbers represent the upper bound
on information that can be associated with any
quantum wave function description of the
electron or the universe.

But this information arises only upon interaction
with observers, prior to which interaction it remains
latent. Because of the unitary nature of the
transformations in a quantum system, information will
manifest itself only to an observer who reduces
the wave function. This means that we must, to remain
consistent, postulate a reduction affected by 
the environment outside of the universe.

\section{Collapse of Wave Function Reconsidered}

Let us consider a quantum system consisting
of a large number of particles. To the observer outside
the system, the particles, being indistinguishable, are
simultaneously present at all the locations.
Or we can represent the particles as a wave which cannot be
localized.

Let us now shift the frame of reference to the quantum
system. Within it, particles do not have a point
of reference. Since each particle is equivalent
to any other, each particle may be supposed to
be located at the same point. 
This may sound strange but it is no more
so than particles flying backward in time
or universe splitting into many copies of itself.
It means that with respect to an observer on 
a particle, space has collapsed so that the
locations of particles inferred by an observer
outside actually belong to the same position.
In other words,
the particle space will appear very different
from classical space to an observer within.

Now let us imagine that this quantum system
interacts with an observer outside of the system.
This will cause the wave function of the
system to collapse
into one
of the component states. The particles, which were
simultaneously present all over the space (with respect
to the outside observer), now find
themselves in definite locations. From the
point of an observer within the system, the 
particles which were earlier co-located now fly
off extremely rapidly to acquire definite locations.
Or the single wave now spawns many particles.

This process is akin to a sudden expansion
of the space, as the objects occupy definite possibilities
in the classical space of the observer.
In this view, ordinary space arises after the
quantum system has decohered. Subsequent to
this collapse, one is justified in
the use of classical notions of space.

When this system has interacted with an outside one the
system decoheres.
Each particle sees its
neighbor travel away to a distance that, on average, corresponds
the dimensions of the decohered system.
In other words, the observer within perceives
the system to expand very quickly.

This expansion may be seen to be in two phases:
the initial phase of
rapid expansion, representing the
inflationary phase; a slower later phase, arising
out of the slowed later decoherence.

If photons are transmitted by particles that are undergoing
decoherence, it will appear to an observer outside of
the system that they were traveling at speeds much faster
than the speed of light.
This increase in the apparent speed could, in principle,
be measured.
This expansion may be seen as a mutually repulsive
force amongst the particles.

If the system is large, then one must also consider the
scenario of the observer system interacting with it in a 
graduated manner. This case will be taken
up in the next section.

\section{Expansion of Space in Cosmology}

We assume that the ideas applicable to individual quantum systems
are also applicable to the universe. Corresponding to
the instantaneous reduction of the wave function for individual
quantum systems is the inflation of the universe.

Let's consider that
at the initial condition the universe 
consisted of $n$ particles in the
same state. Consider, further, that Big Bang 
is a result of the interaction of the universe $U$
with the observer $M$, causing the state to
decohere. The $n$ particles would then
get distributed over
the $N$ states.

It is realistic to assume that the 
interaction of $U$ and $M$ occurs somewhat gradually.
We assume that $M$ is sufficiently 
inhomogeneous so that we can look at its interaction
with $U$ as a sequence of interactions mediated
through the subsystems $M_i$.  
This does two things: i) it makes the overall
evolution of $U+M$ nonlinear, ii) it causes
an immediate reduction of parts of the wave function of
$U$, followed by further continuing reduction in a
recursive manner.

The wave function $| Cosmos \rangle $ of the universe
could then be written as:

\[ | Cosmos\rangle = \sum_i a_i | U_i \rangle | M_{1i} \rangle \]
when the first subsystem $M_1$ has interacted with $U$.
This is followed by interactions of the interpenetrating
$U+M_i$ with other parts of $M_i$.

The interaction with $M_1$ will lead to a rapid expansion.
As other $M_i$s interact, we will witness expansion in
smaller clusters. This process will act in a recursive fashion
so that the regions initially separated in large clumps will
be reduced in the next step, and so on.

The reduction by this mechanism ensures that the
universe will have homogeneity because $U$ is initially in a 
superposition of all states.

\subsection*{Decoherence times}

Joos and Zeh [5] have calculated the effect of
decoherence from an external environment. The speed
of decoherence depends on the number of
particles of the observer $M$ interacting with the system. The
decoherence factor $g (t)$ is calculated to be

\[ |g (t) | \approx e^{- q t } \]
where $q$ is the total number of particles
interacting with the system during the time $t$. Since this exponent
is a large quantity, the decoherence will take
place extremely rapidly.

Consider the estimates of
Joos and Zeh as a guide, where objects
(ranging from dust to big molecule) under various
conditions, from vacuum on earth in full light of
the sun to laboratory vacuum, take from
$10^{-13}$ to $10^{-36}$ seconds to decohere.
(These conditions, rather than that of
intergalactic vacuum, which is a consequence
of the reduction, appear reasonable from the point
of view of a comparison.)
In a more direct interaction of $U$ and $M$, the decoherence times
would be much smaller.
The details of the interaction between $U$ and $M$ will
determine the actual values of the decoherence.
But one would expect that as the two systems begin their interpenetration,
the expansion accelerates to a peak and then 
becomes progressively smaller in magnitude.
This is clear by considering the two extreme cases:
i) before the interaction begins, there is no expansion;
ii) once the interaction is complete, the expansion stops.

It is interesting that the inflationary phase
of the universe is estimated to have lasted from
$10^{-35}$ to $10^{-24}$ seconds, an interval which sits
around the values obtained for conditions on earth.

In practical terms, to an observer situated
within, it would appear that
the system expanded almost instantaneously. 
If it is assumed that the parts of the
system kept in touch by means of photons, then
it would be necessary to postulate that the
speed of light in this phase was much greater
than what it is now.

\section{Variable Speed of Light}

It has been suggested that accepting that
the speed of light $c$ in the early universe varied
and it was much higher provides resolution to the
horizon and flatness problems of cosmology [6].
In the standard Big Bang model, the homogeneity,
isotropy, and flatness of the universe are features
that are a result of the beginning as initial
conditions.
On the other hand,
variable speed of light makes it possible for
the photons to have had the time to smooth out
the temperature and density irregularities in
the parts which have never interacted with each
other.
But this proposal and its variants require
acceptance of time change of other fundamental
constants. Ideas related to the time change
of electric charge, $\hbar$ and the gravitational
constant have been variously examined [7].
Therefore, this idea, although helpful in some
sense, raises other difficulties.

In our view of expansion of space
as a consequence of the collapse of
the wave function of the universe, there
is no need to consider these further 
time-varying constants of nature.
The variation in the speed of light is just a
projection of the collapse of the universe's
wave function to the reference frame of an
observer within the system.

The observations from within 
the system that show the particles flying off
instantaneously can be seen within the
framework of a superluminal expansion or, if
one wished to preserve the speed of light as
the ultimate limit, by postulating 
that the speed of light was much greater at
epochs close to Big Bang and thereafter it became
smaller.
But this variable speed of light is only
an artifact of the model of explanation.

The characteristics of the early universe could
be used to infer properties of $M$ and the nature
of interaction of $U$ and $M$.

\section{Discussion}

The collapse scenarios sketched here are 
admittedly rather vague; they are intended just
as an outline of a new way of looking at
a quantum system from within.
But these ideas have testable implications. One
would expect to see higher speed of light
for photons originating from atoms in a quantum system
that is undergoing decoherence. But this higher
speed of light would merely be an artifact related
to the observer on the atom.

Our ideas exhibit scale invariance
and this permits some testing of cosmological
models in the laboratory without the need for any
new physics.
This is in contrast to recent theories where
exotic forms of matter and forces have been
postulated to account for anomalous
observations at the cosmic level.

The view of expansion of space as a result of
the collapse of the wave function of the universe
provides us with the possibilities of new
experiments to check these ideas. 

The idea  of observation of the
wave function of the universe raises many questions. How
did the interaction between the universe $U$ and the
environment $M$ take place?
How did the giant wave function of $U$ get formed at
the initial condition?
Is there an infinity of universes that lie beyond
ours?

\end{document}